\documentclass{PoS}
\usepackage[english]{babel}
\usepackage[utf8]{inputenc}
\usepackage{cite}
\usepackage{xspace}
\usepackage[T1]{fontenc} 
\usepackage{subfig}
\usepackage{tabularx}  
\usepackage{longtable}  %
\usepackage{booktabs} %
\usepackage{units} %
\usepackage{comment}
\usepackage{pict2e}
\def\LV{\ifmmode {\mathrm{LIV}}\else{\scshape LIV}\fi\xspace}
\def\LIV{\ifmmode{\mathrm{LIV}}\else{\scshape LIV}\fi\xspace}
\def\LI{\ifmmode {\mathrm{LI}}\else{\scshape LI}\fi\xspace}
\def\Ec{\ifmmode {\mathrm{E_c}}\else{\scshape $\rm E_c$}\fi\xspace}
\def\hEc{\ifmmode {\mathrm{\hat{E}_c}}\else{\scshape $\rm \hat{E}_c$}\fi\xspace}
\def\ELIV{\ifmmode {\mathrm{E_{\LIV}}}\else{\scshape $\rm E_{\LIV}$}\fi\xspace}
\def\ELIVl{\ifmmode {\mathrm{E_{\LIV}^{(1)}}}\else{\scshape $\rm E_{\LIV}^{(1)}$}\fi\xspace}

\title{The Spectrum of the Crab Nebula and Highest Energy Photons Measured by HAWC}

\ShortTitle{HAWC Crab Spectrum and VHE Photons}

\author{      
    \speaker{J. T. Linnemann}$^{\rm a}$
    J. P. Harding$^{\rm b}$
    J. Lundeen$^{\rm a,b}$
    S. Marinelli$^{\rm a}$ 
    and H. Martínez-Huerta$^{\rm c}$
    \ \ \ \ \ \ \ \ \ \ \ \ \ \ \ \ \ \ \ \ \ \ \ \ \ \ 
    for the HAWC Collaboration\footnote{For a complete author list, see http://www.hawc-observatory.org/collaboration/icrc2019.php} \\ \\
    $^{\rm a}$ Department of Physics and Astronomy, Michigan State University, MI, USA \\ 
    $^{\rm b}$ Physics Division, Los Alamos National Laboratory, Los Alamos, NM, USA \\
    $^{\rm c}$ Instituto de Física de São Carlos, Universidade de São Paulo, São Carlos, SP, Brasil \\

        E-mail: \email{linnemann@pa.msu.edu}}

\abstract{
HAWC has developed new energy algorithms using an artificial neural network for event-by-event reconstruction of Very High Energy (VHE) primary gamma ray energies. Unlike previous estimation methods for HAWC photons, these estimate photon energies with good energy precision and accuracy in a range from 1 TeV to greater than 100 TeV. Photon emission at the highest energies is of interest in understanding acceleration mechanisms of astrophysical sources and where the acceleration might cut off. We apply the new HAWC reconstruction to present the preliminary measurement of the highest energies at which photons are emitted by the Crab Nebula and by six additional sources in the galactic plane which emit above 50 TeV. We have observed photons above 200 TeV at 95\% confidence. We also compare fits to the HAWC Crab spectrum with other measurements and theoretical models of the Crab spectrum.}

\FullConference{36th International Cosmic Ray Conference -ICRC2019-\\
		July 24th - August 1st, 2019\\
		Madison, WI, U.S.A.}

\begin{document}


\section{Introduction}
The High Altitude Water Cherenkov  Observatory (HAWC) is located at 4100 m altitude at 19º N latitude near the Sierra Negra volcano in Puebla, México.   The HAWC array consists of 300 tanks each containing 200,000 liters of water over an area of 22,000 m$^2$.  Each tank contains 4 photomultipler tubes to detect Cherenkov radiation from air shower particles entering the water.   The detector configuration allows a wide instantaneous field of view, simultaneously viewing all sources within 45º of zenith.  HAWC has operated since 2015 with $> 95$\% up-time.  

HAWC  recently reported\cite{HAWC_CRAB_2019} on the development of two new energy estimation techniques which allow for event-by-event measurement of TeV photons, and their application to the measurement of the spectrum of the Crab Nebula to above 100 TeV.  In this contribution we first describe the Crab spectrum results, and compare various fits to the Crab spectrum.  Then we apply the new energy estimation  to the analysis of the highest energy photons in the spectrum of the Crab, and of other newly-identified sources of photon emission above 50 TeV.

Here we briefly summarize the performance of the two energy estimators; more details are given in \cite{HAWC_CRAB_2019}.   The Ground Parameter (GP) energy estimator is based on the fit of measured events to a lateral distribution shape, evaluated at an optimal radius, and corrected for zenith angle.   The Neural Network (NN) energy estimator combines various measures of event size, the zenith angle, the location of the shower on the array, and the lateral distribution of the shower represented in fractional charge deposited in rings about the shower location.   The performance of the estimators based on shower simulations is given in Fig.~\ref{figs:rms}(a), which shows the resolution and bias of each energy estimator added in quadrature.  Above 10 TeV, the RMS approaches .1 in log$_{10}$ space, or a bit above 30\% in linear energy space.  This performance is a significant improvement over the original energy estimation technique used in HAWC\cite{hawc_crab_2017}, and allows us to measure the spectrum of sources to 100 TeV and beyond.

\section {Crab Spectrum}
\begin{figure}[h!]
    \centering
    \subfloat[]{{\includegraphics[height=.38\linewidth]{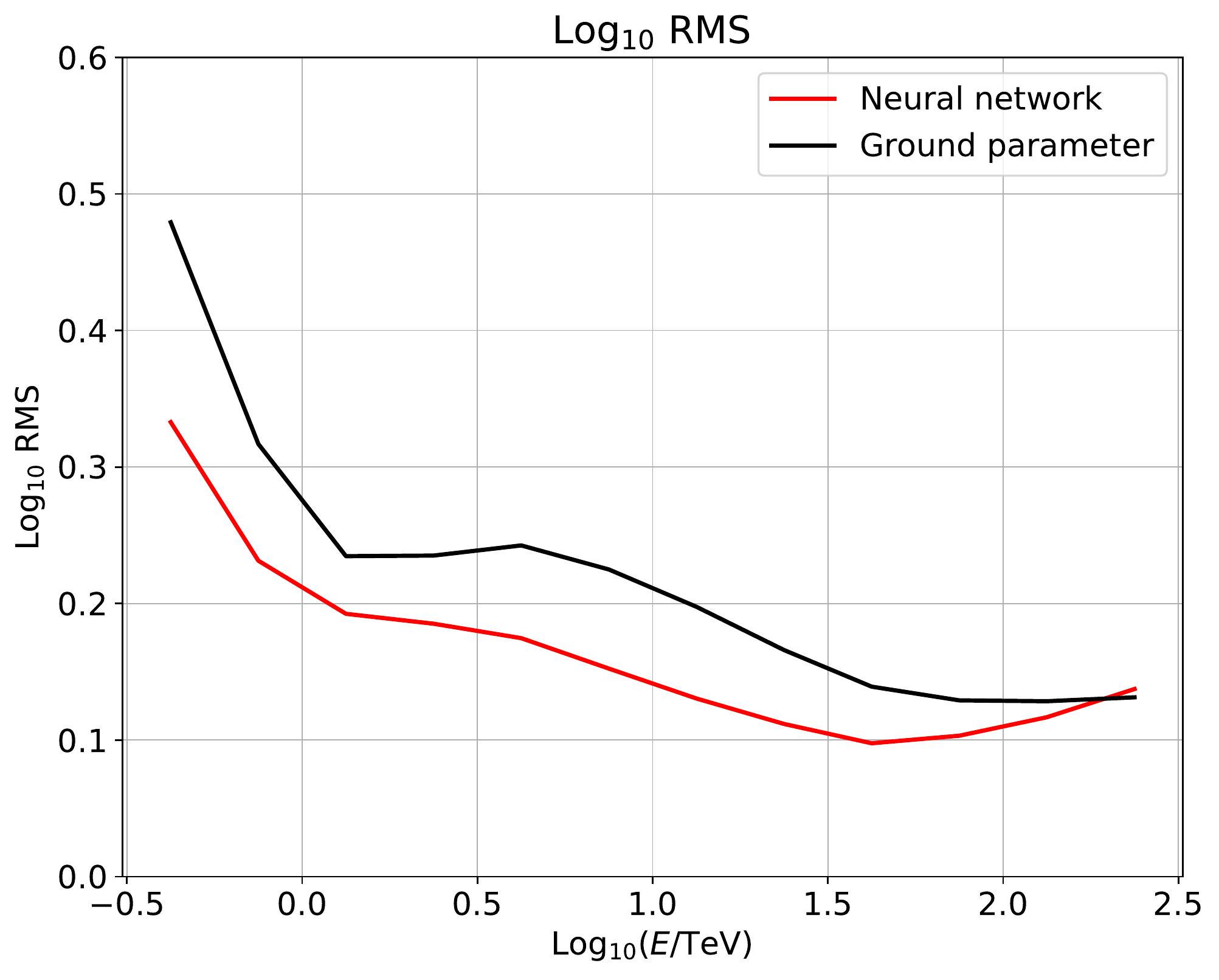} }}%
    \subfloat[]{{\includegraphics[height=.39\linewidth]{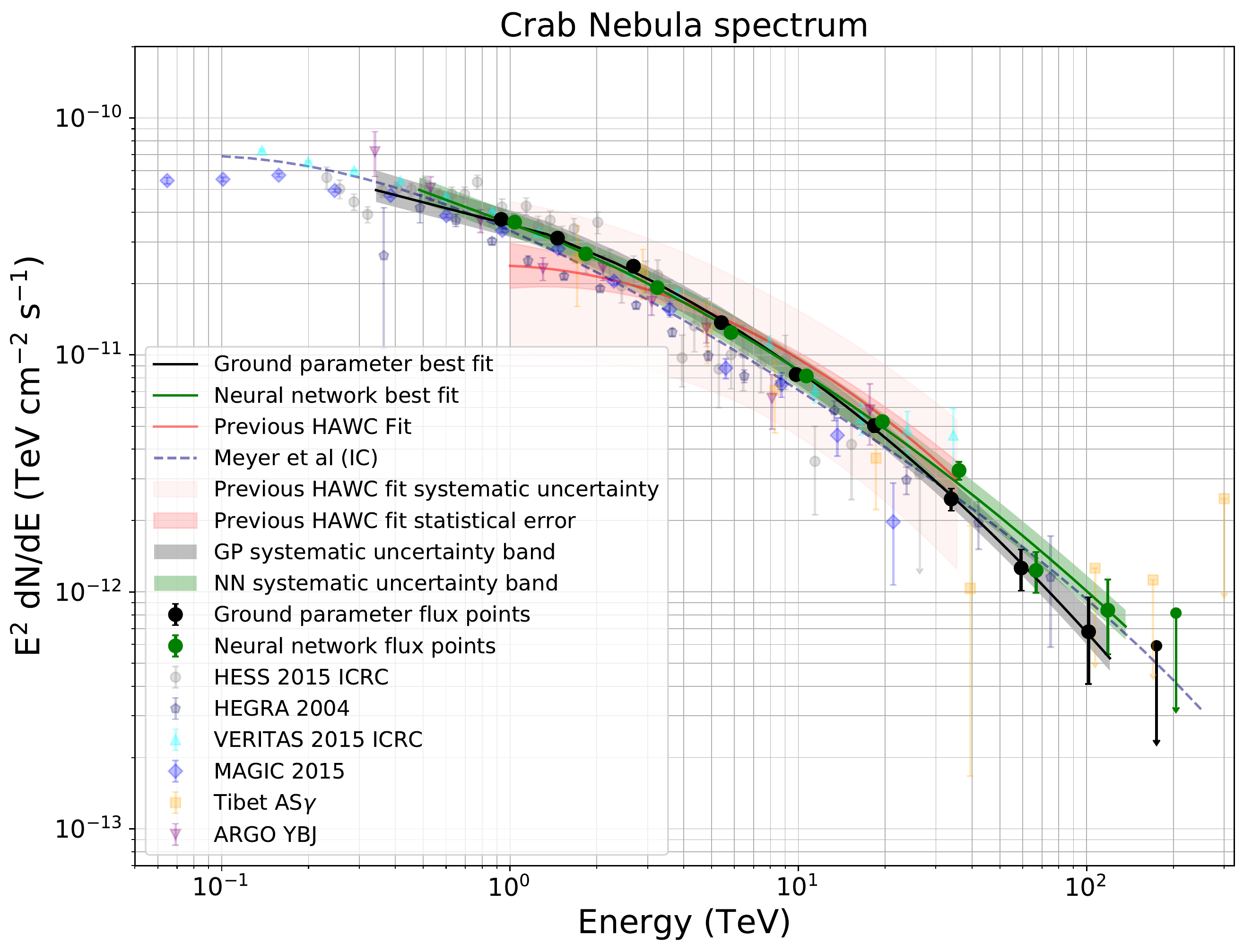} }}%
    \caption{ Left (a).  The RMS in Log space of the two energy estimators.  This is defined as $RMS \equiv \sqrt{\left\langle\left(\log_{10} \hat{E} - \log_{10}E\right)^2 \right\rangle}.$\ \ \        Right (b).The HAWC Crab Spectrum.
    }   
    \label{figs:rms}%
\end{figure}

The analysis of the spectrum uses forward folding of a fit spectrum to data binned in two variables: $fhit$, the fraction of the array PMTs which have shower hits, and $\hat{E}$, the energy estimator.  The $fhit$ binning allows better modeling of the point spread function than using just the energy estimator alone.  Fig.~\ref{figs:rms}(b) shows the resulting spectrum of the Crab Nebula.   The colored bands give the systematic error of the HAWC spectra for each of the estimators.  Ref.~\cite{HAWC_CRAB_2019} discusses HAWC's study of systematic error effects in detail. The two spectra are in strong agreement within systematics except at the highest energy, where they are still in agreement within statistical errors.  Also shown in the figure are the previous HAWC fit and data points from a number of other VHE observatories\cite{hess_2015,veritas_2015,magic_crab_2015,Tibet_2015,Argo_2015,HEGRA_2004}.

\section {Fits to Crab Spectrum}

\begin{figure}[h!]
    \centering
    \includegraphics[width=\textwidth]{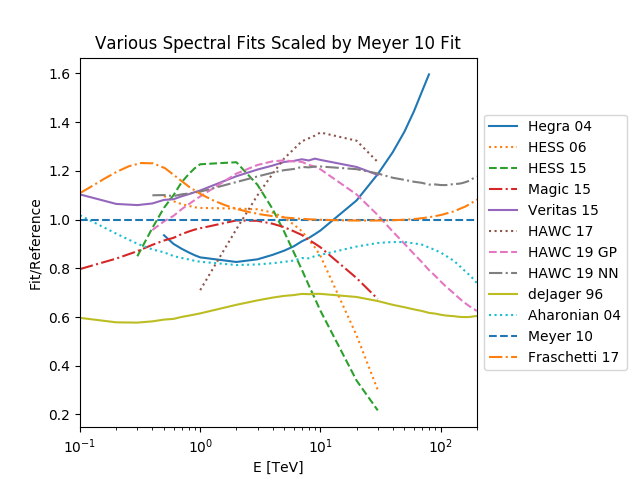}
    \caption{ Here are ratios of various fits to dN/dE, divided (for reference) by the Inverse Compton fit from Meyer~\cite{Meyer2010}. The other theory fits include de Jager~\cite{DeJager1996}, Aharonian~\cite{HEGRA_2004}, and Fraschetti~\cite{Fraschetti2017}.  Experimental references include HAWC~\cite{hawc_crab_2017,HAWC_CRAB_2019}, HEGRA~\cite{HEGRA_2004}, HESS~\cite{hess_2006,hess_2015}, MAGIC~\cite{Magic_2016}, and VERITAS~\cite{veritas_2015}. 
    }
    \label{figs:fits}%
\end{figure}

In Fig.~\ref{figs:fits} we compare a number of fits to the VHE spectrum. First we consider several theoretical calculations.  We have normalized all the fits to the Inverse Compton (IC) model calculated in Meyer~\cite{Meyer2010}.  We note that this work also attempted to harmonize the energy scales of experiments available as of 2010.  This calculation considers a constant magnetic field model of the Crab, with two electron populations and five sources of target photons, and fits 10 parameters to data ranging from the synchrotron to TeV.  Other IC models agree within a factor of two, with more recent models producing more flux.  The Fraschetti and Pohl model~\cite{Fraschetti2017} considers a single electron population with a log-parabola momentum distribution, plus synchrotron self-Compton and and cosmic microwave background as target photons, and produces similar results to Meyer above 1 TeV, but differs at lower energies.

Next we consider various experiments' fits to their own VHE data.  The recent HAWC data extend to higher energies than the previous HAWC data; the new HAWC data lies somewhat above predictions and other measurements above 10 TeV. However, other than HEGRA, most data stopped before 40 TeV, and the fits tended to drop rapidly above 40 TeV.  Most experiments fit their data with a log-parabola form, but this  curved spectrum is not a fundamental description, and the fit parameters depend significantly on the energy range used in the fit.   It is also worth mentioning that because data uncertainty varies with energy, fits may emphasize the data in different portions of the experimental energy range: comparing fits is useful, but does not fully replace looking at the actual data points.  HAWC data suggests the spectrum continues beyond 10 TeV more strongly than implied by HESS or MAGIC fits.  In contrast, the HEGRA fit uses a simple power law, which at highest energy has less curvature than HAWC observes and IC predicts. 

This simple comparison does not consider the systematic errors on energy scales, and suggests a further study with newer data to that which was carried out in Meyer\cite{Meyer2010}.  On the other hand, most measurements are compatible at the 20\% level in flux, at least in the range 1 to 10 TeV.


\section{Analysis of High Energy Photons}

In this section we study the highest energy photons in the Crab spectrum and in other Galactic sources seen by HAWC.
Since the emphasis here is on the upper extremes of the spectrum, a number of details are changed compared to the analysis of the Crab spectrum.   
First we concentrate on the NN energy estimator as it is expected to have better energy resolution~(see Fig. \ref{figs:rms}(a) ).  Second, we re-bin the two highest bins of estimated energy, subdividing both the $(56,100)$ and the $(100,177)$ TeV bins into three finer bins each.      

We consider, besides the Crab, six other sources\cite{HAWC_ICRC19_Kelly} which have evidence of emission at high energy:  although the Crab is the brightest VHE source, other sources with a harder spectrum might be sources of higher-energy photons.  The analysis needs a 
spectrum model for each source.   The Crab is modeled 
with a log-parabola spectrum.   The other six sources, shown in Table~\ref{tabs:tbl}, are Galactic sources all within 2 degrees of the Galactic plane.  They are all modeled 
with spectra taken as power laws with an exponential cutoff.  Finally, to desensitize us to imperfect modeling of the point spread function, the analysis is carried out in tophat bins chosen for each source to be large enough that the results no longer depend on the choice of tophat radius.
\begin{figure}[h]
    \centering
    \subfloat[]{{\includegraphics[height=.32\linewidth]{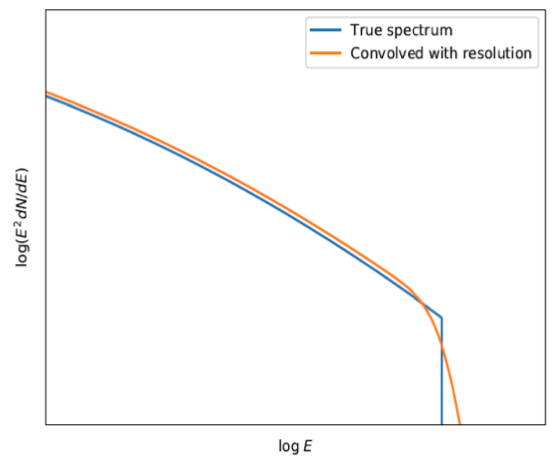} }}%
    \subfloat[]{{\includegraphics[width=.48\linewidth]{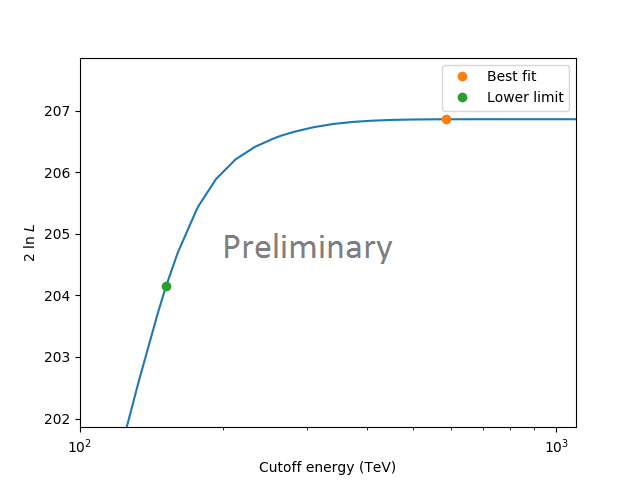} }}%
    \caption{ Left (a). True spectrum with hard cutoff at some energy $\rm E_c$ and the expected observed spectrum due to the detector energy resolution~\cite{Sam_Thesis}. Right (b). Likelihood curve as a function of the Energy cutoff in the Crab analysis; the lower point (green) shows the lower limit at 95$\%$ CL.
    }
    \label{figs:cut_like}%
\end{figure}

This analysis provides a lower limit for an energy $E_c$ beyond which there is weak or no evidence for continuation of emission for a spectrum.\footnote{The same method is also used for searching for violation of Lorentz Invariance, which \textit{does} involve a sharp endpoint to the spectrum. This is described in Refs. \cite{HAWC_ICRC19_Humberto,Sam_Thesis,HAWC_LIV_CPT}.  A future publication will also consider the effects of systematic errors.}   The principle is illustrated in Fig.~\ref{figs:cut_like}.    We perform a fit to the chosen energy spectrum shape, and compare the fit likelihood with an energy spectrum with a hard cutoff at energy $E_c$ as shown in Fig.~\ref{figs:cut_like} (a).  The hard cutoff model is not astrophysically motivated, but it has the virtue of being sensitive to whether the spectrum above $E_c$ is inconsistent with (has more photons than) the background above $E_c$ plus the expected amount of mis-reconstructed events with energy $E\ <\ E_c$ due to the spectrum below $E_c$. Thus this test is independent of the actual spectral shape above $E_c$.  Specifically, we plot the profile likelihood (with spectral fit parameters optimized for each $E_c$) as a function of $E_c$.

First we consider  whether sources show actual preference for such a hard cutoff. The statistical test is to calculate the log likelihood ratio of the fit with no cutoff and the fit including such a cutoff.  
\begin{equation}
D = 2\ln \left(\frac{\mathcal{L}(\hEc)}{\mathcal{L}(\hEc \rightarrow \infty)}\right).
\end{equation}
We calculate the p-value corresponding to finding a value of $D$ being the observed value or larger, if the true spectrum has no hard cutoff.   The p-value should be distributed uniformly between 0 and 1.  The p-values shown in the Table\ref{tabs:tbl} indicate that none of the sources prefer a cutoff with even a modest significance level (say less than .05).   Thus assured that our spectra do not indicate a preference for $E_c < \infty$, we proceed to set a lower limit on $E_c$,

The change in $\mathcal{L}$ compared to the best fit value can be mapped into the size of a confidence interval.   We consider here two intervals, $95\%$ and $99.73\%$ (`` $3\sigma$ '').  From the shape of the likelihood curve in Fig~\ref{figs:cut_like} (b), it is clear that this limit is intrinsically one-sided, as we lose statistical power to identify a finite $E_c$ for large values of $E_c$.  The corresponding values of $2\ \Delta \ln \mathcal{L}$ for the intervals are $2.71, 7.74$.

\begin{table}[h]
\centering
{\begin{tabular}{@{}ccccccc@{}}\toprule
Source & $E_c(95\%)$ & $E_c(3\sigma)$ & p-value \\
&TeV &TeV&\\
\toprule
2HWC J1825-134 & 253 & 168  & 1.000\\
2HWC J1908+063 & 213 & 156 & 0.990\\
Crab (HAWC) & 152 & 96  & 1.000\\
2HWC J2031+415 & 144 & 78 & 0.714\\
2HWC J2019+367 & 121 & 86  & 0.8282\\
J1839-057 & 79 & 66  & 0.357\\
2HWC J1844-032 & 77 & 63  & 0.294\\
\end{tabular}
}
\caption{PRELIMINARY: HAWC Sources and Photon Energy Limits.}
\label{tabs:tbl}
\end{table}

We see from Table~\ref{tabs:tbl} that at $95\%$ CL, HAWC has seen $>  100$ TeV photons from five sources, and at $3 \sigma$ equivalent, saw $> 100$ TeV photons from two sources.

\section{Conclusions}

HAWC has measured the Crab spectrum to beyond 100 TeV with  new estimators which have good energy resolution out to 100 TeV and beyond.  The resulting spectrum  is in agreement within 20\% in flux of measurements from IACTs, where there are overlapping measurements.  We see that the two most recent IC calculations agree above 1 TeV within 10\% of each other, and up to about 100 TeV, HAWC measurements are within 20\% of these predictions.  We have also examined the spectra of the Crab and six other Galactic sources for high-energy photon production (above background and bin migration from mis-measured lower-energy photons), and found 95\% CL evidence for 100 TeV photons from five sources, and $3\sigma$ evidence from two sources above 100 TeV.

\newpage


\begin{thebibliography}{10}
\expandafter\ifx\csname url\endcsname\relax
  \def\url#1{{\tt #1}}\fi
\expandafter\ifx\csname urlprefix\endcsname\relax\def\urlprefix{URL }\fi
\providecommand{\eprint}[2][]{\url{#2}}

\bibitem{HAWC_CRAB_2019}
Abeysekara A~U {\em et~al.\/} (HAWC) 2019  (\textit{Preprint}
  \eprint{1905.12518})

\bibitem{hawc_crab_2017}
Abeysekara A~U {\em et~al.\/} 2017 {\em The Astrophysical Journal\/} {\bf 843}
  39
  \urlprefix\url{http://iopscience.iop.org/article/10.3847/1538-4357/aa7555/meta}

\bibitem{hess_2015}
Holler M {\em et~al.\/} 2015 {\em Proceedings of Science (34th International
  Cosmic Ray Conference)\/}  236 (\textit{Preprint} \eprint{1509.02902})

\bibitem{veritas_2015}
Meagher K 2015 {\em Proceedings of Science (34th International Cosmic Ray
  Conference)\/} {\bf 236} ISSN 18248039

\bibitem{magic_crab_2015}
Aleksi{\'{c}} J {\em et~al.\/} 2015 {\em Journal of High Energy Astrophysics\/}
  {\bf 5-6} 30--38 ISSN 22144048 (\textit{Preprint} \eprint{1406.6892})

\bibitem{Tibet_2015}
Amenomori M {\em et~al.\/} 2015 {\em Astrophys. J.\/} {\bf 813}

\bibitem{Argo_2015}
Bartoli B {\em et~al.\/} {\em The Astrophysical Journal\/}  119 ISSN 1538-4357

\bibitem{HEGRA_2004}
Aharonian F~A {\em et~al.\/} 2004 {\em Astrophys. J.\/} {\bf 614} 897--913 ISSN
  0004-637X (\textit{Preprint} \eprint{0407118})

\bibitem{Meyer2010}
Meyer M, Horns D and Zechlin H 2010 {\em Astronomy {\&} Astrophysics\/} {\bf
  A2}

\bibitem{DeJager1996}
de~Jager O~C, Harding A~K, Michelson P~F, Nel H~I, Nolan P~L, Sreekumar P and
  Thompson D~J 1996 {\em The Astrophysical Journal\/} {\bf 457} 253 ISSN
  $\backslash$
  \urlprefix\url{http://labs.adsabs.harvard.edu/adsabs/abs/1996ApJ...457..253D/}

\bibitem{Fraschetti2017}
{Fraschetti} F and {Pohl} M 2017 {\em Mon. Not. Roy. Astron. Soc} {\bf 471} 4856--4864
  (\textit{Preprint} \eprint{1702.00816})

\bibitem{hess_2006}
{Aharonian} F, {Akhperjanian} A~G, {Bazer-Bachi} A~R, {Beilicke} M, {Benbow} W,
  {Berge} D, {Bernl{\"o}hr} K, {Boisson} C, {Bolz} O and {Borrel} V 2006 {\em
  Astron. Astrophys} {\bf 457} 899--915 (\textit{Preprint} \eprint{astro-ph/0607333})

\bibitem{Magic_2016}
Aleksic J {\em et~al.\/} 2016 {\em Astroparticle Physics\/} {\bf 72} 76--94

\bibitem{HAWC_ICRC19_Kelly}
Malone K (HAWC) 2019 {First HAWC spectra of Galactic gamma-ray sources above
  100 TeV and the implications for cosmic-ray acceleration} {\em Proceedings,
  36th International Cosmic Ray Conference (ICRC2019): Madison, WI, U.S.A.,
  July 24th - August 1st, 2019.\/}  (\textit{Preprint}
  \eprint{1908.07059})

\bibitem{Sam_Thesis}
Marinelli S 2019 {PhD Thesis, Michigan State University}
  {$hawc-observatory.org/publications/ \#thesis$}

\bibitem{HAWC_ICRC19_Humberto}
Martínez-Huerta H {\em et~al.\/} (HAWC) 2019 {Constraints on Lorentz
  Invariance Violation Using HAWC Observations above 100 TeV} {\em Proceedings,
  36th International Cosmic Ray Conference (ICRC2019): Madison, WI, U.S.A.,
  July 24th - August 1st, 2019.\/}  (\textit{Preprint}
  \eprint{1908.09614})

\bibitem{HAWC_LIV_CPT}
{J T Linnemann for the HAWC Collaboration} (HAWC) 2019 {Lorentz Invariance
  Violation Limits from HAWC } {\em {8th Meeting on CPT and Lorentz Symmetry
  (CPT'19) Bloomington, Indiana, USA, May 12-16, 2019}\/} (\textit{Preprint}
  \eprint{--})

\end{thebibliography}

\providecommand{\newblock}{}

\end{document}